\documentclass[12pt]{article}
\usepackage[a4paper,left=2cm,right=1.2cm,top=3cm,bottom=4cm]
{geometry}
\usepackage{amsmath,amstext, amsthm, empheq, extpfeil,  
amsbsy,amssymb,marvosym,fancyhdr,graphicx,amscd,amsfonts,latexsym,delarray,stackrel,lineno,color,cite,appendix,xcolor,url,hyperref}
\usepackage{wasysym}
\usepackage{subfig}

\usepackage{srcltx}
\usepackage{mathrsfs}
\usepackage{float} 
\usepackage[all]{xy}
\usepackage{t1enc}
\usepackage{mathrsfs}
\usepackage{pifont}
\usepackage{authblk}

\usepackage{mathpple}

\usepackage[T1]{fontenc}

\definecolor{Green}{rgb}{0,1,0}
\definecolor{Blue}{RGB}{0,0,191}
\definecolor{mathmodecolor}{RGB}{0,102,0}
\definecolor{keywordcolor}{RGB}{0,51,151}
\definecolor{sourcebackgroundcolor}{RGB}{255,247,223}
\definecolor{unixagred}{RGB}{255,0,0}
\definecolor{lightgray}{RGB}{191,191,191}
\definecolor{green}{RGB}{1,191,191}

\newcommand*\patchAmsMathEnvironmentForLineno[1]{%
  \expandafter\let\csname old#1\expandafter\endcsname\csname #1\endcsname
  \expandafter\let\csname oldend#1\expandafter\endcsname\csname end#1\endcsname
  \renewenvironment{#1}%
     {\linenomath\csname old#1\endcsname}%
     {\csname oldend#1\endcsname\endlinenomath}}%
\newcommand*\patchBothAmsMathEnvironmentsForLineno[1]{%
  \patchAmsMathEnvironmentForLineno{#1}%
  \patchAmsMathEnvironmentForLineno{#1*}}%
\AtBeginDocument{%
\patchBothAmsMathEnvironmentsForLineno{equation}%
\patchBothAmsMathEnvironmentsForLineno{align}%
\patchBothAmsMathEnvironmentsForLineno{flalign}%
\patchBothAmsMathEnvironmentsForLineno{alignat}%
\patchBothAmsMathEnvironmentsForLineno{gather}%
\patchBothAmsMathEnvironmentsForLineno{multline}%
}

\newtheorem{thm}{Theorem}[section]
\newtheorem{prop}[thm]{Proposition}
\newtheorem{cor}[thm]{Corollary}
\newtheorem{lem}[thm]{Lemma}

\newtheorem{defn}[thm]{Definition}
\newtheorem{rem}[thm]{Remark}

\def\Aut{{\rm Aut}}

\def\id{{\rm id}}

\def\Trace{{\rm Tr}}
\def\Tr{{\rm Tr}}

\def\C{{\mathbb C}}

\def\R{{\mathbb R}}
\def\Z{{\mathbb Z}}

\def\Tr{{\rm Tr}}

\def\cA{{\mathcal A}}

\def\cC{{\mathcal C}}

\def\cE{{\mathcal E}}

\def\cH{{\mathcal H}}

\def\cL{{\mathcal L}}

\def\dar[#1]{\ar@<2pt>[#1]\ar@<-2pt>[#1]}
\def\qqq{\,,\,~\forall}

\newcommand{\ie}{{\it i.e.\/}\ }

\newcommand{\nil}[1]{}
\newcommand{\noopsort}[1]{}

\makeatletter
\DeclareMathOperator{\exterior}{\@ifnextchar^\@exterior{\@exterior^{}}}
\def\@exterior^#1{\mathop{\bigwedge\nolimits^{\!#1}}}
\makeatother

\title{Entropy and the spectral action}
\author
      {Ali H. Chamseddine$^{1,3,5}$, Alain Connes$^{2,3,4}$ and Walter D. van
Suijlekom$^{5}$
      \\
     $^{1}$Physics Department, American University of Beirut, Lebanon\\
            $^{2}$College de France, 3 rue Ulm, F75005, Paris, France\\
      $^{3}$I.H.E.S. F-91440 Bures-sur-Yvette, France\\
      $^{4}$Department of Mathematics, Ohio State University, Columbus OH 43210 USA\\
      $^{5}$Institute for Mathematics, Astrophysics and Particle Physics, Radboud
University Nijmegen, Heyendaalseweg 135, 6525 AJ Nijmegen, The Netherlands.
\\
\bigskip
\texttt{chams@aub.edu.lb, alain@connes.org, waltervs@math.ru.nl}
      }

\begin{document}

\maketitle
\begin{abstract}
We compute the information theoretic von Neumann entropy of the state associated to the fermionic second quantization of a spectral triple. We show that this entropy is given by the spectral action of the spectral triple for a specific universal function. The main result of our paper is the surprising relation between this function and the Riemann zeta function. It manifests itself in particular by the values of the coefficients $c(d)$ by which it multiplies the $d$ dimensional terms in the heat expansion of the spectral triple. We find that $c(d)$ is the product of the Riemann xi function evaluated at $-d$ by an elementary expression. In particular $c(4)$ is a rational multiple of $\zeta(5)$ and $c(2)$ a rational multiple of $\zeta(3)$. The functional equation gives a duality between the coefficients in positive dimension, which govern the high energy expansion, and the coefficients in negative dimension, exchanging even dimension with odd dimension.
\end{abstract}
\tableofcontents
\section{Introduction}
The spectral action principle \cite{AC2} has so far been developed for arbitrary  test functions of the form $\chi(D^2/\Lambda^2)$ of a spectral triple $(\cA,\cH,D)$.
The spectral action defined, for a fixed choice of $\chi$, as 
$$
\Tr(\chi(D^2/\Lambda^2))
$$
is an additive functional of the spectral triple $(\cA,\cH,D)$, \ie when evaluated on the direct sum of two spectral triples it behaves additively. This additivity is, at the conceptual level, a key property of the spectral action. Motivated by  the formalism of Quantum field theory, one gets another natural way to obtain 
an additive functional of spectral triples: given $(\cA,\cH,D)$ one performs the fermionic second quantization using the Clifford algebra of the underlying real Hilbert space $\cH_\R$ and evaluating the  von Neumann information theoretic entropy of the unique  state satisfying the KMS condition at inverse temperature $\beta$ with respect to the time evolution of the Clifford algebra induced by the operator $D$. The second quantization transforms direct sums into tensor products and moreover the von Neumann entropy of the tensor product of two states is the sum of the respective entropies. Thus one indeed obtains an additive functional of spectral triples  in this way, which we simply call the ``entropy'' $S(\cA,\cH,D)$ of the spectral triple. After recalling  the KMS condition in \S \ref{sect:KMS} and the fermionic second quantization in \S \ref{sectfermion} we show in \S \ref{sectentropy} that the entropy functional is  the spectral action for a specific universal test function $\chi(x)=h(\sqrt x)$. The main result of this paper is the surprising intimate relation, explored in \S \ref{secth} between the function $h$ and the Riemann zeta function.  This is explained at the conceptual level to try to understand why it appears.  

The content of this paper is purely mathematical. As far as physical applications are concerned, using the role of the spectral action as a framework to unify all fundamental interactions, this work indicates in particular the need to develop the spectral approach in connection with second quantization.

\subsection*{Acknowledgements}
The work of A. H. C is supported in part by the National Science Foundation
Grant No. Phys-1518371. A. H. C would also like to thank Institute for Mathematics, Astrophysics and Particle Physics, Radboud University Nijmegen for hospitality where part of this work was done. The involved research has partly been enabled by a Radboud Excellence Professorship awarded to Prof. Chamseddine. 

WvS would like to thank Sijbrand de Jong for stimulating discussions on entropy and the spectral action. WvS thanks IH\'ES for hospitality during a visit in early 2018, as well as NWO for support via VIDI-grant 016.133.326.

\section{Spectral triples and second quantization}
We use the operator algebra formalism of $C^*$-dynamical systems and KMS condition to pass from the first-quantized or ``one-particle'' level of spectral triples $(\cA,\cH,D)$ to the second-quantized level.  The Hilbert space $\cH$ is used to construct the complexified Clifford algebra $C:={\rm Cliff}_\C(\cH_\R)$ of its underlying real Hilbert space $\cH_\R$ when considered as a Euclidean space. The operator $D$ is used as the generator of a one-parameter group $\sigma_t\in \Aut(C)$ of automorphisms of the Clifford algebra. The algebra $\cA$ manifests itself through the semigroup of inner fluctuations \cite{CCS13} which labels deformations $D'$ of the operator $D$. The action of the semigroup of inner fluctuations continues to make sense at the second-quantized level and gives rise to deformations  $\sigma'_t\in \Aut(C)$ of the above one-parameter group of automorphisms.  
 We concentrate here on the meaning of the spectral action, and thus take the $C^*$-dynamical system $(\cC,\sigma_t)$ as our starting point, keeping in mind that the results will automatically apply to the deformations $(\cC,\sigma'_t)$.
 \subsection{KMS and the dynamical system $({\rm Cliff}_\C(\cH_\R)$, $\sigma_t={\rm Cliff}(e^{itD}))$}\label{sect:KMS}
We first briefly recall the KMS condition for $C^*$-dynamical systems, that is, a
$C^*$-algebra $\cC$ together with a one-parameter group of
automorphisms $\sigma_t\in \Aut(\cC)$, $t\in \R$.
\begin{defn}\label{KMScondDef}
Let  $(\cC,\sigma_t)$ be a $C^*$-dynamical system.  For a given
$0<\beta<\infty$, a state $\varphi$ on the unital $C^*$-algebra
$\cC$ satisfies the KMS condition at inverse
temperature $\beta$ if
 for all $a,b \in \cC$, there exists a function 
$F_{a,b}(z)$ which is holomorphic on the strip
\begin{equation}\label{KMSstrip}
I_\beta =\{ z\in \C\,  |\, 0< \Im(z) <\beta \},
\end{equation}
continuous on the boundary $\partial I_\beta$ and bounded, with the
property that for all $t\in \R$
\begin{equation}\label{KMScond}
F_{a,b}(t)=\varphi(a\sigma_t(b)) \ \ \text { and } \ \
F_{a,b}(t+i\beta)=\varphi(\sigma_t(b)a).
\end{equation}
\end{defn}
In short the KMS condition at inverse
temperature $\beta$ means that one has the formal equality 
\begin{equation}\label{KMScond1}
\varphi(a\sigma_t(b))\vert_{t=i\beta}=\varphi(ba)\qqq a,b\in \cC.
\end{equation}
What matters in the context of the present paper is  the existence and uniqueness of KMS states on a matrix algebra $M_n(\C)$ and we give the short proof for convenience. A one-parameter group of automorphisms $\sigma_t\in \Aut(M_n(\C))$ is always associated to a self-adjoint $H=H^*\in M_n(\C)$ by  
$$
\sigma_t(A)= e^{itH}Ae^{-itH}\qqq t\in \R,   ~ A\in M_n(\C).
$$
 Given a state $\psi$ on the matrix algebra $M_n(\C)$ there exists a unique density matrix $\rho\geq 0$ such that 
$$
\psi(T)=\Tr(\rho T) \qqq T \in M_n(\C).
$$
By uniqueness of the trace it follows that 
$$
\psi(AB)=\psi(BA) \qqq A,B\Rightarrow \rho = \frac 1n \, \id.
$$
A state which is KMS$_\beta$ for $\sigma_t\in \Aut(M_n(\C))$ is invariant. In fact 
$$
\psi(e^{-\beta H} Be^{\beta H})=\psi(B) \qqq B\Rightarrow  e^{\beta H}\rho e^{-\beta H}=\rho
\Rightarrow \rho e^{\beta H}=e^{\beta H}\rho.
$$
 It follows using \eqref{KMScond1} that if
$\psi$ is KMS$_\beta$ then, with $B'=e^{-\beta H} B$
$$
\psi(BA)=\psi(Ae^{-\beta H} Be^{\beta H})\Rightarrow \Tr(\rho e^{\beta H}B'A)= \Tr(\rho A B' e^{\beta H})
$$
so that $\rho e^{\beta H}=e^{\beta H}\rho$ defines a trace and hence is a scalar multiple of $\id$. This shows that $\rho=Z e^{-\beta H}$ for $Z=1/\Tr(e^{-\beta H})$ and gives the uniqueness of the KMS$_\beta$ state. The same formula gives the existence. 
For completeness, we include a proof of the following result on KMS-states on Clifford algebras ({\em cf.} \cite[Prop. 5.2.23]{BR97}).
\begin{prop}\label{KMS}
Let $\cH$ be a complex Hilbert space, $D$ a self-adjoint operator in $\cH$ with compact resolvent. Let $C:={\rm Cliff}_\C(\cH_\R)$	be the complexified Clifford algebra of the underlying real Hilbert space $\cH_\R$ and $\sigma_t\in \Aut(C)$  be the one-parameter group of automorphisms associated to $\exp(itD)\in \Aut(\cH_\R)$. Then for any $\beta>0$ there exists a unique KMS$_\beta$ state $\psi_\beta$ on the $C^*$-dynamical system $(C,\sigma_t)$. 
\end{prop}
\proof  One applies the existence and uniqueness of KMS states on a matrix algebra to the subalgebra of $C={\rm Cliff}_\C(\cH_\R)$ associated to the subspace corresponding to a finite dimensional spectral projection of $D$. This is enough to prove the uniqueness of the KMS$_\beta$ state. The existence also follows since the existence part for matrix algebras gives a coherent system of states which define a state on the inductive limit of the $C^*$-algebras. \endproof 
\begin{prop}\label{KMS1}
Let $\cH$, $D$,  $C:={\rm Cliff}_\C(\cH_\R)$, $\sigma_t\in \Aut(C)$ and $\psi_\beta$	 be as in Proposition \ref{KMS}. Then if the operator $\exp(-\beta\vert D\vert)$ is of trace class, the state $\psi_\beta$ is of type I and the associated irreducible representation is given by the Fermionic second quantization associated to the complex structure $I:=i \ {\rm sign}\, D$ on $\cH_\R$. 
\end{prop}
The proof of this proposition will be given in \S \ref{sectfermion}, Proposition \ref{KMS2} after recalling some terminology. 

\subsection{Fermionic second quantization}\label{sectfermion}
In this section we recall the procedure of (fermionic) second quantization. We refer to \cite{Ara87,CarR87}  and \cite[Sect. 5.3 and 6.1]{GVF01} for excellent expositions on fermionic second quantization.

Consider the real Euclidean vector space $V := \cH_\R$ that underlies the complex Hilbert space $\cH$. Our first goal is to find irreducible representations of the Clifford algebra associated to the real Euclidean vector space $\cH_\R$ underlying $\cH$ and for this it turns out that a crucial role is played by complex structures.  We let $I$ be an orthogonal complex structure on $V$, which is not necessarily the one coming from $\cH$. Then we may regard $V$ as a complex vector space when we define $i$ to act as $I$. The resulting complex Hilbert space will be denoted by $V_I$. 

A representation of the complexified Clifford algebra ${\rm Cliff}_\C(V)$ is given on the Fock space $\exterior V_I$ that is built on $V_I$ by the usual formula
\begin{align*}
\gamma_I : {\rm Cliff}_\C(V)& \to \cL (\exterior V_I)\\
v &\mapsto a_I^*(v) + a_I(v) ;\qquad  (v \in V).
\end{align*}
Here the {\em creation operators} $a_I^*(v)$ depend $\C$-linearly on $v \in V_I$ and are given by exterior multiplication by $v$ while the {\em annihilation operator} $a(v)$ is its adjoint. We choose a unit vector $\Omega_I \in \exterior^0 V_I$ and call it the {\em vacuum vector}. It is annihilated by $a_I(v)$ for all $v \in V$. The following is well-known.
\begin{lem}
The above representation $\gamma_I$ of the complexified Clifford algebra ${\rm Cliff}_\C(V)$ on Fock space $\cL (\exterior V_I)$ is irreducible. 
\end{lem}
\proof
We may assume that $V$ is a inductive limit of finite-dimensional Hilbert spaces, and, accordingly, that ${\rm Cliff}_\C(V)$ is the $C^*$-algebraic inductive limit of finite-dimensional Clifford algebras. Without loss of generality we may thus assume that $\dim V < \infty$ so that ${\rm Cliff}_\C(V)$ are simple matrix algebras. We invoke Schur's Lemma to conclude that $\gamma_I$ is irreducible if and only if every operator $T:V \to V$ commuting with all $\gamma_I(v)$ ($v \in V^\C$) is a scalar. 

For any $v$, one has, using the $\C$-linearity of $a_I^*(v)$ and $\C$-anti-linearity of $a_I(v)$ 
$$
 a_I^*(v)=\frac 12\left(\gamma_I(v)-i\gamma_I(Iv) \right), \ \  a_I(v)=\frac 12\left(\gamma_I(v)+i\gamma_I(Iv) \right).$$
Hence any $T$ that commutes with $\gamma_I(v)$ for all $v$ commutes with $a_I^*(v)$ and $a_I(v)$. From this it follows that
$$
a_I(v) (T \Omega_I) = T (a_I(v)\Omega_I)= 0 
$$
so that $T \Omega_I \in \Lambda^0(V_I)$. In other words, $T \Omega_I = t \Omega_I$ for some $t \in \C$. Moreover, 
$$
T(v_1 \wedge \cdots \wedge v_k) =  T \left( a_I^*(v_1) \cdots a_I^* (v_k) \Omega \right) = \left(a_I^*(v_1) \cdots a_I^* (v_k) \right) (T \Omega) =  t \left(v_1 \wedge \cdots \wedge v_k\right),
$$
so that $T = t \cdot {\rm id}$. 
\endproof

Any orthogonal operator $T:V \to V$ induces an automorphism of ${\rm Cliff}_\C(V)$ by sending $\gamma_I(v) \to \gamma_I(Tv)$. In some cases this automorphism can be lifted to the Fock space $\exterior V_I$, for instance, if $U$ is a unitary operator on $V_I$. Then, if $\exterior U$ is the unitary operator in the Fock space such that on simple tensors
$$
\exterior U(v_1\wedge \cdots \wedge v_n):=U(v_1)\wedge \cdots \wedge U(v_n),
$$
one has the covariance 
$$
\exterior U\circ a_I^*(v)\circ \exterior U^*= a_I^*(Uv).
$$
We thus get the equality 
$$
(\exterior U) \gamma_I(v) (\exterior U^*)=\gamma_I(Uv).
$$
\begin{rem}
More generally speaking, the {\em Shale--Stinespring Theorem} \cite{SS65} says that the automorphism on ${\rm Cliff}_\C(V)$ defined by an orthogonal operator $T: V \to V$ is implementable by a unitary operator on Fock space $\exterior V_I$ if and only if $T + I T I $ is Hilbert-Schmidt.  
\end{rem}

Suppose now that we are given a (complex) Hilbert space $\cH$ and a self-adjoint operator $D$ in $\cH$ with compact resolvent. Again, let $V= \cH_\R$ denote the underlying real vector space. Suppose that we take the natural complex structure on $V$ so that $V_I = \cH$. Then the above construction gives us an irreducible representation $\gamma$ of the canonical anti-commutation relations (CAR) algebra on $\exterior \cH$ but, from a physical point of view this representation is not the right one to consider. In fact, Dirac realized in \cite{Dir27} that in order to avoid unwanted negative energy solution to his Dirac equation, one has to fill up (what is now called) the {\em Dirac sea}. Mathematically speaking, this construction is nicely captured by choosing another irreducible representation, corresponding to a different complex structure on $\cH$. Let us describe it in some detail.

If $E_\pm$ are the spectral projections of $D$ corresponding to the positive and negative eigenspaces of $D$ in $\cH$, let us define the following complex structure:
$$
I= i (E_+ - E_-). 
$$ 
In other words, $I = i F$ where $F= D |D|^{-1}$ is the sign\footnote{We take the convention that the sign of $0$ is $1$} of $D$. In view of the previous section, this gives rise to another irreducible representation $\gamma_I$ of ${\rm Cliff}_\C(V)$ in Fock space, where the key difference with respect to the original Fock space representation $\gamma$ is that $i$ acts as $-i$ on $E_-( \cH)$. In other words, the operator $D$ can be considered to act as $|D|$ which now only has positive eigenvalues. More precisely,
\begin{prop}\label{KMS2}
$(i)$~The one-parameter group $\sigma_t\in \Aut(C)$ is implemented in the (physical) Fock representation  by the one-parameter unitary group $W(t)=\exterior \exp(it |D|)$, \ie one has  
\begin{equation}\label{dirac1}
\gamma_I(\sigma_t(A))=\exterior(e^{it |D|})\gamma_I(A)\exterior(e^{-it |D|})\qqq A \in {\rm Cliff}_\C(V).
\end{equation}

$(ii)$~If  $\exp(-\beta |D|)$ is of trace class  the state $\psi_\beta$ is of type I
and is given by 
\begin{equation}\label{dirac2}
\psi_\beta(A)=\frac 1Z\Tr\left(\exterior \exp(-\beta |D|) \gamma_I(A)\right)\qqq A \in {\rm Cliff}_\C(V)
\end{equation}
where the normalization factor $Z$ is finite.
\end{prop}
\proof The associated KMS$_\beta$ state is obtained after normalization from the density matrix $W(i\beta)$ obtained by analytic continuation. Thus it coincides here with the operator 
$$
\rho=\exterior \exp(-\beta |D|).
$$
Since $T=\exp(-\beta |D|)$ is positive and of trace class we get that $\rho=\exterior T$ is also positive and of trace class (with trace given by the determinant of $1+T$).  Thus $Z<\infty$. The state $\psi_\beta$ is implemented by a density matrix in an irreducible representation and is thus of type I.\endproof

\section{von Neumann information theoretic entropy}\label{sectentropy}
We start by briefly recalling von Neumann's notion of entropy. Consider a density matrix $\rho$ on a Hilbert space $\cH$, {\it i.e.} a positive trace-class operator with normalized trace. It induces a state $\phi$ on any $C^*$-subalgebra of $\cL(\cH)$ by setting $\phi( \cdot ) = \Tr ( \rho \cdot)$. The {\em entropy} of this state $\phi$ is then defined to be
$$
S(\phi) := - \Tr (\rho \log \rho).
$$
For composite systems $\phi_1 \otimes \phi_2$ on $\cH_1 \otimes \cH_2$ one finds the following important additivity property for entropy
$$
S(\phi_1 \otimes \phi_2 ) = S(\phi_1)  + S(\phi_2).
$$

Let us start with a basic example of entropy that will play a crucial role in what follows. 

\begin{lem}\label{functionE} Let $x>0$, the entropy of the partition of the unit interval in two intervals with ratio of size $x$, is given by 
$$
\cE(x):=\log (x+1)-\frac{x \log (x)}{x+1}.
$$	
\end{lem}
\proof The sizes of the intervals are $\frac{1}{x+1}$ and $\frac{x}{x+1}$. One has 
$$
-\frac{\log \left(\frac{1}{x+1}\right)}{x+1}-\frac{x \log \left(\frac{x}{x+1}\right)}{x+1}=\frac{x+1}{x+1}\log (x+1)-\frac{x \log (x)}{x+1}=\cE(x).  
$$
\endproof
\begin{cor} 
\label{corl:E-inverse}
One has $\cE(x)=\cE(1/x)$ for any $x>0$.	
\end{cor}
\proof 
The obtained partitions are isomorphic. 
\endproof 
The following result gives an expression for the entropy of density matrices that arise as a second-quantized operator. 
\begin{lem}
\label{lma:entropy-trace-E}
 Let $T\in \cL^1(\cH)^+$ be a positive trace class operator and $\phi$ the state associated to $\exterior T$ then 
\begin{equation}\label{entropstate}
	S(\phi)=\Trace(\cE(T)).
\end{equation}	
\end{lem}
\proof Let first $T=T_1\oplus T_2$ be an orthogonal decomposition. Let us show that the associated states fulfill $\phi=\phi_1\otimes \phi_2$. One has with $\rho_I=\exterior T_I$ the equality $\rho=\rho_1\otimes \rho_2$ by the compatibility of the wedge functor with direct sums. Then 
$$
\Trace(\rho_1\otimes \rho_2)=\Trace(\rho_1)\Trace(\rho_2)\Rightarrow \phi=\phi_1\otimes \phi_2.
$$
Next the entropy functional fulfills 
$$
S(\phi_1\otimes \phi_2)=S(\phi_1)  +S(\phi_2).
$$
This shows that the functional $S(\phi)$ is additive for direct sum decompositions. It also applies to infinite sums and one can thus consider only the one dimensional case. In this case the state associated to the operator $T$ of multiplication by $x$ corresponds to $\exterior T$ whose spectrum is $\{1,x\}$ and hence has entropy given by the function $\cE(x)$ of Lemma \ref{functionE}.\endproof

\begin{thm}\label{KMS4}
Let $\cH$, $D$,  $C:={\rm Cliff}_\C(\cH_\R)$, $\sigma_t\in \Aut(C)$ and $\psi_\beta$ be as in Proposition \ref{KMS}. Then if the operator $\exp(-\beta\vert D\vert)$ is of trace class, the state $\psi_\beta$ is of type I and its von Neumann entropy is 
equal to the spectral action $\Tr(h(\beta D))$ for the spectral function $h(x):=\cE(e^{-x})$.
\end{thm} 
\proof  
 The first statement follows from Proposition \ref{KMS2}. The statement about the entropy follows from Lemma \ref{lma:entropy-trace-E} together with the fact that $x \mapsto \cE(e^{-x})$ is an even function ({\em cf.} Corollary \ref{corl:E-inverse}).
\endproof

\section{The function $h$ and Riemann's $\xi$ function}\label{secth}

Theorem \ref{KMS4} provides a very specific choice of test function for the spectral action and in this section we shall analyze the surprising relation between this specific function and the Riemann $\xi$ function. Thus consider
\begin{align*}
h(x)&:=\cE(e^{-x})= \frac{x}{1+e^x} + \log (1+e^{-x})\\
&= \log (2)-\frac{x^2}{8}+\frac{x^4}{64}-\frac{x^6}{576}+\frac{17 x^8}{92160}-\frac{31 x^{10}}{1612800}+O\left(x^{11}\right).
\end{align*}

\begin{lem}\label{functionh} The function $h(x)$ is an even positive function of the variable $x\in \R$. Its derivative is 
$$
h'(x)=- \frac{x}{4 \cosh ^2\left(\frac{x}{2}\right)}.
$$
\end{lem}
\begin{figure}[H]
\begin{center}
\includegraphics[scale=0.6]{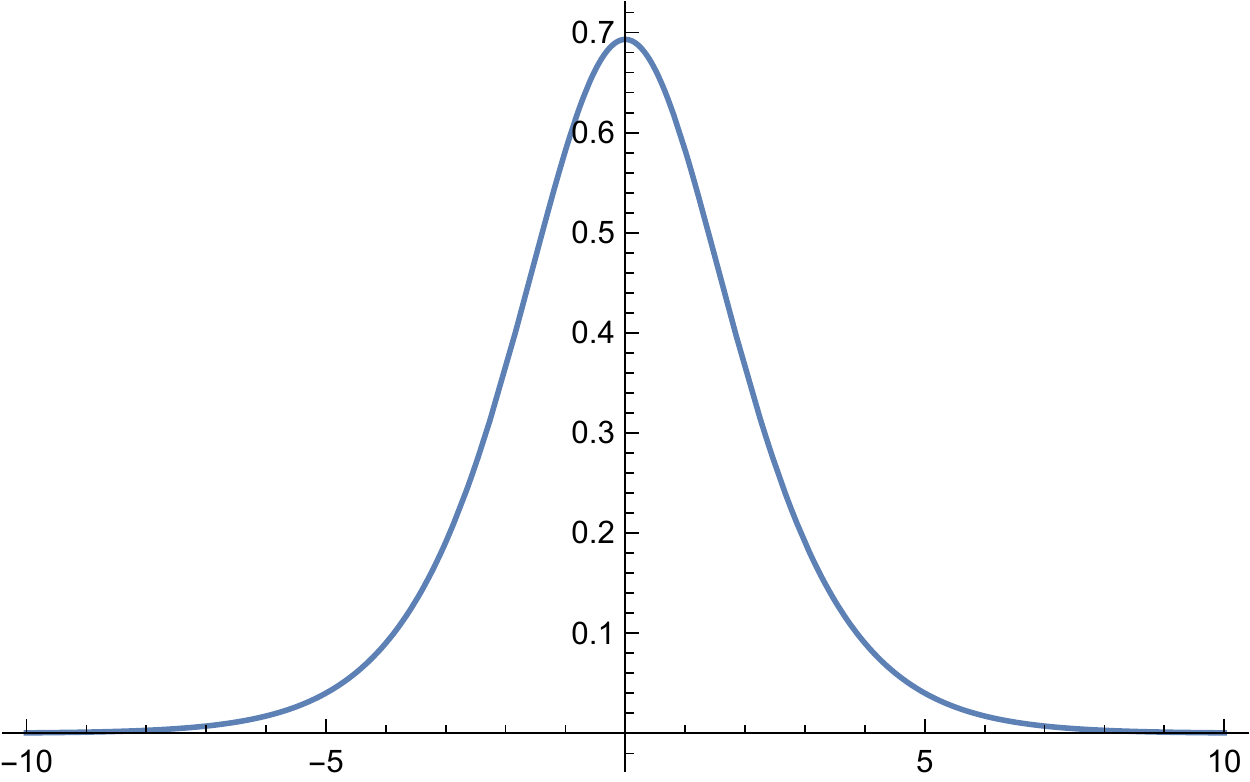}
\end{center}
\caption{Graph of the function $h$. \label{graphofh} }
\end{figure}

\begin{lem}\label{functionhprime} One has, for $x\geq 0$,  
$$
\sum_\Z \frac{(2 \pi  n+\pi )^2-x}{\left((2 \pi  n+\pi )^2+x\right)^2}= \frac{1}{4 \cosh ^2\left(\frac{\sqrt x}{2}\right)}.
$$
\end{lem}
\proof The simplest periodic function written as a convergent Eisenstein series is
\begin{equation}\label{Eisenstein}
	\sum_\Z \frac{1}{(\pi  n+x)^2}=\frac{1}{(\sin \,x)^2}.
\end{equation}
One has 
$$
\frac{(2 \pi  n+\pi )^2+u^2}{\left((2 \pi  n+\pi )^2-u^2\right)^2}=\frac{1}{2 (-2 \pi  n+u-\pi )^2}+\frac{1}{2 (2 \pi  n+u+\pi )^2}.
$$
Thus if we take $u=i\sqrt x$ we obtain 
$$
\sum_\Z \frac{(2 \pi  n+\pi )^2-x}{\left((2 \pi  n+\pi )^2+x\right)^2}=
\sum_\Z \frac{1}{ ( 2\pi  n+u+\pi )^2}=\frac 14 \sum_\Z \frac{1}{ ( \pi  n+u/2+\pi/2 )^2}
$$
which, using \eqref{Eisenstein}, gives the required equality since $\sin (u/2+\pi/2 )=\cos(u/2)=\cosh \sqrt x/2$.\endproof 
Next one has 
$$
\int_0^\infty \left(2 \pi ^2 (2 n+1)^2 t-1\right) e^{\pi ^2 \left(-(2 n+1)^2\right) t-t x}dt=\frac{(2 \pi  n+\pi )^2-x}{\left((2 \pi  n+\pi )^2+x\right)^2}.
$$
We want to apply the Fubini theorem to justify the formula
\begin{equation}\label{Fubini}
	\frac{1}{4 \cosh ^2\left(\frac{\sqrt x}{2}\right)}=\int_0^\infty g(t)e^{-t x} dt, \ \ g(t):=\sum_\Z \left(2 \pi ^2 (2 n+1)^2 t-1\right) e^{\pi ^2 \left(-(2 n+1)^2\right) t}.
\end{equation}
It suffices to show that the double sum of absolute values is convergent, \ie
$$
\int_0^\infty \sum_\Z \vert 2 \pi ^2 (2 n+1)^2 t-1 \vert e^{\pi ^2 \left(-(2 n+1)^2\right) t-t x}dt<\infty .
$$
One has $v e^{-v}\leq e^{-v/2}$ and $e^{-v}\leq e^{-v/2}$ for $v>0$ so that the above integral is bounded by
$$
2\int_0^\infty \sum_\Z  e^{\pi ^2 \left(-(2 n+1)^2\right) t/2-t x}dt 
$$ 
which is finite since for $t$ small $\sum_\Z  e^{\pi ^2 \left(-(2 n+1)^2\right) t/2}=O(\frac{1}{\sqrt t})$ while it decays exponentially for large $t$. This justifies the formula \eqref{Fubini} but we now need a better estimate of the function $g(t)$ when $t\to 0$. 
\begin{lem}\label{thetafunction} One has   
\begin{equation}\label{Poissonsum}
	g(t)=\frac{-1}{4 \sqrt{\pi } t^{3/2}}\sum_\Z  (-1)^n n^2 q^{n^2}, \ \ q=e^{-\frac{1}{4 t}}.
\end{equation}
\end{lem}
\proof We apply the Poisson summation formula, for the Fourier transform 
$$
F(f)(\omega):=\int_{-\infty }^{\infty } f(u) e^{2 i \pi  u \omega } \, dt
$$
and the function 
$$
f(u):=\left(2 \pi ^2 (2 u+1)^2 t-1\right) e^{\pi ^2 \left(-(2 u+1)^2\right)  t}.
$$
One has
$$
F(f)(\omega)=-\frac{ \omega ^2 e^{-i \pi \omega}e^{-\frac{ \omega^2  }{4 t}}}{4 \sqrt{\pi } t^{3/2}}
$$
Thus \eqref{Poissonsum} follows from \eqref{Fubini} and the Poisson summation formula 
$\sum_\Z f(n)=\sum_\Z F(f)(m)$. \endproof 

This formula \eqref{Poissonsum}  together with the lack of constant term in the function $\sum_\Z  (-1)^n n^2 q^{n^2}=q\partial_q\theta_4(0,q)$ show that the function $\tilde g(t):=g(t)/2t$ is of rapid decay near $t=0$ and in particular is integrable. 
\begin{prop}
One has 
\begin{equation}\label{Poissonsum1}
	h(x)=\int_0^\infty e^{-tx^2}\tilde g(t) dt.
\end{equation}	
\end{prop}
\proof Let $k(x)$ be the right hand side of \eqref{Poissonsum1}. One has, using \eqref{Fubini}, 
$$
\partial_x k(x)= -2x \int_0^\infty e^{-tx^2} t\tilde g(t) dt=-x\int_0^\infty g(t)e^{-t x^2} dt=
	\frac{-x}{4 \cosh ^2\left(\frac{x}{2}\right)}=\partial_x h(x).
$$
Thus the two functions $h,k$ are equal because they both tend to zero when $x\to \infty$. \endproof 
Note that in terms of theta functions one has by Jacobi triple product formula 
$$
\theta_4(0,q)=\sum_\Z  (-1)^n q^{n^2}=\prod_1^\infty(1-q^{2m})(1-q^{2m-1})^2.
$$
This formula gives us a good control of the sign of the function $\sum_\Z  (-1)^n n^2 q^{n^2}=q\partial_q\theta_4(0,q)$. Indeed one has $0<q<1$ so that all terms in the product are positive. Moreover the logarithmic derivative gives 
$$
q\partial_q\theta_4(0,q)=\theta_4(0,q)\sum_1^\infty\left(-\frac{2 m q^{2 m}}{1-q^{2 m}}-\frac{2(2 m-1) q^{2 m-1}}{1-q^{2 m-1}}\right)  \leq 0.
$$
Thus by \eqref{Poissonsum} we get that both $g(t)$ and $\tilde g(t)$ are positive functions. 

We now compute the moments of the function $h$ which are relevant in dimension $4$, they are given by
\begin{lem}\label{momentsofh} $(i)$~One has   
\begin{equation}\label{moments}
	2\int_0^\infty h(x)xdx= \frac{9 \,\zeta (3)}{2}, \ \ 2\int_0^\infty h(x)x^3dx=\frac{225\, \zeta (5)}{4}.
	\end{equation}
	$(ii)$~More generally
	\begin{equation}\label{moments1}
	\int_0^\infty h(x)x^\alpha dx= \frac{1-2^{-\alpha -1}}{\alpha+1} \Gamma(\alpha +3)\zeta(\alpha +2).	\end{equation}
\end{lem} 
\proof One has for $\alpha>0$, integrating by parts, 
\begin{equation}\label{moments2}
\alpha\int_0^\infty h(x)x^{\alpha-1}dx=\int_0^\infty\frac{x^{\alpha+1}}{4 \cosh ^2\left(\frac{x}{2}\right)}dx.
\end{equation}
Moreover by Lemma \ref{functionhprime} one has
$$
\frac{1}{4 \cosh ^2\left(\frac{x}{2}\right)}
=2\sum _{n=0}^{\infty } \frac{(2 \pi  n+\pi )^2-x^2}{\left((2 \pi  n+\pi )^2+x^2\right)^2}.
$$
Consider the function 
$$
f(x):=\frac{x^2-1}{\left(x^2+1\right)^2}.
$$
One has 
$$
x^{-2}f(\frac{2 \pi  n+\pi }{x})=\frac{(2 \pi  n+\pi )^2-x^2}{\left((2 \pi  n+\pi )^2+x^2\right)^2}
$$
while for $\Re(y)\in (-1,1)$
$$
\int_0^\infty f(a)a^y da =\frac{\pi  y}{2  \cos \left(\frac{\pi  y}{2}\right)},
$$
so that 
$$
\int_0^\infty x^{-2}f(\frac{2 \pi  n+\pi }{x})x^{\alpha +1} dx=
\int_0^\infty f(\frac{2 \pi  n+\pi }{x})x^{\alpha -1} dx  
$$
and changing variables by $x'=\frac{2 \pi  n+\pi }{x}$ gives for $\Re(\alpha +1)\in (-1,1)$
$$
\int_0^\infty x^{-2}f(\frac{2 \pi  n+\pi }{x})x^{\alpha +1} dx=(2 \pi  n+\pi )^\alpha 
\int_0^\infty f(x')x'^{-\alpha-1}dx=(2 \pi  n+\pi )^\alpha
\frac{-\pi (1+\alpha)}{2  \cos \left(\frac{-\pi (1+\alpha)  y}{2}\right)}.
$$ 
Thus for $\Re(\alpha )\in (-2,-1)$ we get the equality 
$$
\sum _{n=0}^{\infty }\int_0^\infty x^{-2}f(\frac{2 \pi  n+\pi }{x})x^{\alpha +1} dx=
\frac{-\pi (1+\alpha)}{2  \cos \left(\frac{-\pi (1+\alpha)  }{2}\right)}\pi^\alpha 
\sum _{n=0}^{\infty }(2n+1)^\alpha.
$$
One has 
$$
\sum _{n=0}^{\infty }(2n+1)^\alpha= 2^{\alpha } \left(2^{-\alpha }-1\right) \zeta (-\alpha )
$$
and 
$$
\frac{-\pi (1+\alpha)}{2  \cos \left(\frac{-\pi (1+\alpha)  }{2}\right)}\pi^\alpha=
\frac{\pi^{1+\alpha} (1+\alpha)}{2  \sin \left(\frac{\pi \alpha  }{2}\right)}.
$$
One has the functional equation in the form 
$$
\zeta (1-s)=\frac{2}{(2\pi)^s}\cos \left(\frac{\pi s  }{2}\right)\Gamma(s)\zeta(s)
$$
and it gives for $s=\alpha +1$ 
$$
\frac{\pi^{1+\alpha} (1+\alpha)}{2  \sin \left(\frac{\pi \alpha  }{2}\right)}2^{\alpha } \left(2^{-\alpha }-1\right) \zeta (-\alpha )=\frac 12\left(1-2^{-\alpha }\right)\Gamma(\alpha +2)\zeta(\alpha +1)
$$
and by analytic continuation one gets (in concordance with \cite[Formula 3.527(3)]{GR15})
$$
\int_0^\infty\frac{x^{\alpha+1}}{4 \cosh ^2\left(\frac{x}{2}\right)}dx=
2\sum _{n=0}^{\infty }\int_0^\infty x^{-2}f(\frac{2 \pi  n+\pi }{x})x^{\alpha +1} dx
=\left(1-2^{-\alpha }\right)\Gamma(\alpha +2)\zeta(\alpha +1).
$$ 
Thus using \eqref{moments2} one gets \eqref{moments1}.\endproof 

This gives us for the coefficient of $t^{-a}$ in the heat expansion the term
\begin{equation}\label{moments3}
\frac{1}{\Gamma(a)}\int_0^\infty h(v^{\frac 12})v^{a-1}dv=\frac{2}{\Gamma(a)}
\int_0^\infty h(u)u^{2a-1}du=\frac{(1-2^{-2a})\Gamma(2a+2)\zeta(2a+1)}{a\, \Gamma(a)} .
\end{equation}
We need to check several properties of the right hand side. First it should define an everywhere holomorphic function of $a\in \C$ because as the coefficient of $t^{-a}$ in the heat expansion it is also given, using \eqref{Poissonsum1} by 
$$
\int_0^\infty t^{-a}\tilde g(t) dt
$$
which is an entire function of $a$. Now $\zeta(z)$ has a pole at $z=1$ but the corresponding pole of $\zeta(2a+1)$ is cancelled by the following term which is $0$ at $a=0$: 
$$
\frac{(1-2^{-2a})}{a\, \Gamma(a)}.
$$
The other potential poles come from the term $\Gamma(2a+2)$ which has a pole when $2a+2=-m$ is a negative integer. There are two cases: if $m=2b$ is even, then the term $a\, \Gamma(a)=\Gamma(a+1)$ also has a pole at the negative integer $-b$. If $m=2b+1$ is odd, then the term $a\, \Gamma(a)=\Gamma(a+1)$ does not have a pole at $-b-\frac 12$ but the term $\zeta(2a+1)$ vanishes for $2a+2=-m=-2b-1$ since $2a+1=-2b-2$ is a strictly negative even number. 

Next we should check that when $a$ is a strictly negative integer the formula \eqref{moments3} should give a rational number which agrees up to sign and division by $n!$ with the Taylor expansion of $h(\sqrt{ x})$ at $x=0$. The rationality follows from the rationality of zeta at odd negative integers. 

The expansion of $h(\sqrt{ x})$ at $x=0$ is
$$
h(\sqrt{ x})=\log (2)-\frac{x}{8}+\frac{x^2}{64}-\frac{x^3}{576}+\frac{17 x^4}{92160}-\frac{31 x^{5}}{1612800}+O\left(x^{6}\right).
$$
The terms $c(a):=\frac{(1-2^{-2a})\Gamma(2a+2)}{a\, \Gamma(a)}$ evaluated at negative integers 
are of the form
$$
\text{c}(-1)=-\frac{3}{2},\text{c}(-2)=\frac{15}{4},\text{c}(-3)=-\frac{21}{8},\text{c}(-4)=\frac{17}{16},\text{c}(-5)=-\frac{341}{1120}
$$
while the values of zeta at odd negative integers gives
$$
\zeta (-1)=-\frac{1}{12},\zeta (-3)=\frac{1}{120},\zeta (-5)=-\frac{1}{252},\zeta (-7)=\frac{1}{240},\zeta (-9)=-\frac{1}{132}
$$
and one checks that the Taylor expansion of $h(\sqrt{ x})$ at $x=0$ is given by 
$$
h(\sqrt{ x})=\log (2)+\sum_1^\infty (-1)^n c(-n)\zeta (-2n+1)x^n/n!.
$$
To consider the odd case when $a$ is a negative half integer we use the functional equation to rewrite the terms in \eqref{moments3} as follows
$$
\frac{\Gamma(2a+2)\zeta(2a+1)}{a\, \Gamma(a)}=(2a+1)\frac{\Gamma(2a+1)\zeta(2a+1)}{\Gamma(a+1)}
=\frac{(2a+1)\zeta (-2a)(2\pi)^{2a+1}}{2\cos \left(\pi (a+\frac 12)\right)\Gamma(a+1)}$$ 
$$
=(2a+1)\Gamma(-a)\zeta (-2a)(2\pi)^{2a+1}/(2\pi)
$$
using the formula of complements in the form
$$
\Gamma(-a)\Gamma(a+1)=\frac{\pi}{\cos \left(\pi (a+\frac 12)\right)}.
$$
Moreover in terms of Riemann's $\xi$ function, 
$$
\xi(s):=\frac 12 s(s-1)\pi^{-\frac s2}\Gamma(\frac s2) \zeta(s)
$$
 one has 
$$
(2a+1)\Gamma(-a)\zeta (-2a)(2\pi)^{2a+1}/(2\pi)=2^{2a}\pi^a\xi(-2a)/a
$$
and we get
\begin{equation}\label{moments4}
\frac{(1-2^{-2a})\Gamma(2a+2)\zeta(2a+1)}{a\, \Gamma(a)} =\frac{2^{2a}-1}{a}\pi^a\xi(-2a)
\end{equation}
We thus get, summarizing  the above computations, and changing $a\to -a$,
\begin{lem}\label{xi} The coefficient of $t^{a}$ in the heat expansion is the term
$$
\gamma(a)=\frac{1-2^{-2a}}{a}\pi^{-a}\xi(2a).
$$	
\end{lem}
As a corollary we double check that this is an entire function of the variable $a\in \C$ and that when $a$ is a negative integer $-n$ the result does involve the value of zeta at the odd positive integer $2n+1$. This shows that these odd zeta values show up in the even dimensional case for the coefficients of negative powers of $t$ and in the odd dimensional case for the coefficients of positive powers of $t$. 
$$
\boxed{\begin{aligned}
 \gamma (-2)&=\frac{225 \zeta (5)}{4} \\
 \gamma \left(-\frac{3}{2}\right)&=\frac{14 \pi ^{7/2}}{45} \\
 \gamma (-1)&=\frac{9 \zeta (3)}{2} \\
 \gamma \left(-\frac{1}{2}\right)&=\frac{\pi ^{3/2}}{3} \\
 \gamma (0)&=\log (2) \\
 \gamma \left(\frac{1}{2}\right)&=\frac{1}{2 \sqrt{\pi }} \\
 \gamma (1)&=\frac{1}{8} \\
 \gamma \left(\frac{3}{2}\right)&=\frac{7 \zeta (3)}{8 \pi ^{5/2}} \\
 \gamma (2)&=\frac{1}{32} \\
 \gamma \left(\frac{5}{2}\right)&=\frac{93 \zeta (5)}{32 \pi ^{9/2}} \\
\end{aligned}}
$$ 

What this table of values also shows is that  the functional equation gives a duality between the coefficients of the high energy expansion in even dimension with the coefficients of the low energy expansion in the odd dimensional case.
\appendix
\section{Appendix}

We recall briefly in this appendix the relation of the constants appearing in the heat expansion and the spectral action. Consider a test function $\chi(u)$ which is given as a simple
superposition of exponentials as a Laplace transform of the form
\begin{equation}\label{superpos}
\chi(u) = \int_0^{\infty} e^{-su} \, g(s) \, ds ,
\end{equation}
where the function $g(s) $ is of rapid decay near $0$ and $\infty$.
Let $\Delta$ be a positive self-adjoint operator 
\begin{equation}\label{formal}
\chi(t\Delta) = \int_0^{\infty} e^{-st\Delta} \, g(s) \, ds.
\end{equation}
An asymptotic  expansion of the form ${\rm Trace} \, (\exp(-t\Delta))\sim \sum \, a_{\alpha}\, t^{\alpha}$ then yields
\begin{equation}\label{formal1}
{\rm Trace} \, (\chi (t\Delta)) \sim \sum \, a_{\alpha} \, t^{\alpha}
\int_0^{\infty} s^{\alpha} \, g(s) \, ds \, .
\end{equation}

For $\alpha < 0$ one has, $$s^{\alpha} = \frac{1}{\Gamma (-\alpha)}
\int_0^{\infty} e^{-sv} \, v^{-\alpha -1} \, dv$$ and
$$\int_0^{\infty} s^{\alpha} \, g(s) \, ds = \frac{1}{\Gamma
(-\alpha)} \int_0^{\infty} \chi(v) \, v^{-\alpha -1} \, dv.$$
Moreover for $\alpha=0$ one has 
$$
\int_0^{\infty}  \, g(s) \, ds =\chi(0).
$$
Thus one gets 
\begin{eqnarray}
{\rm Trace} \, (\chi(t\Delta)) &\sim &\sum_{\alpha < 0} a_{\alpha}\; t^{\alpha} \frac{1}{\Gamma
(-\alpha)}\int_0^{\infty} \chi(v) \,
v^{-\alpha -1} \, dv  \nonumber \\
&+ &a_0 \, \chi(0) + \ldots \label{formal2}
\end{eqnarray}
For $\alpha=n$ a positive integer,   using $
(\partial_u)^{n}(e^{-su})=(-1)^n s^n e^{-su}
$, one has\footnote{Note that in the odd-dimensional case $\alpha$ takes half-integral values and this formula does not apply.}
$$
\int_0^{\infty} s^{\alpha} \, g(s) \, ds=(-1)^n\left(\int_0^{\infty} (\partial_u)^{n}(e^{-su}) \, g(s) \, ds\right)_{u=0}=(-1)^n (\partial_u)^{n}  \chi(u)\vert_{u=0}=(-1)^n\chi^{(n)}(0).
$$
In dimension $4$ the values of negative $\alpha$ which are relevant are $-1$ and $-2$, the  coefficients are 
$$
\left(\int_0^{\infty} \chi(v) \, v \, dv\right) t^{-2}a_{-2}+\left(\int_0^{\infty} \chi(v)  \, dv\right) t^{-1}a_{-1} +\chi(0)a_0+\ldots  
$$
and they can be written in terms of  $f(u)=\chi(u^2)$ using
$$
\int_0^{\infty} \chi(v) \, v^{-\alpha -1} \, dv
=2\,\int_{0}^{\infty}f(u)\,u^{-2\alpha-1}\,du .
$$

\end{document}